# オープン IoT に向けた GPU 自動オフロード技術の検討

山登庸次† 出水達也† 野口博史† 片岡操†

† NTT ネットワークサービスシステム研究所，東京都武蔵野市緑町 3-9-11
E-mail: †yamato.yoji@lab.ntt.co.jp

**あらまし** 近年，IoT 技術が進展している．多彩なサービスを実現するために，デバイスとサービスを水平分離的に相互に利活用する，オープン IoT が重視されてきている．私達は，オープン IoT に向け，ユーザが必要なデータを持つデバイスをオンデマンドに発見し，利用する，Tacit Computing 技術を提案している．しかし，既存の Tacit Computing は、性能や運用コストを度外視していた．本稿では，性能改善を行うため，遺伝的アルゴリズムを用いて，GPU 自動オフロードを行う Tacit Computing の要素技術を提案する．GPU オフロードの有効性を確認するため，C/C++アプリケーションの行列計算を評価し，35 倍以上の性能向上を 1 時間以内の探索時間で確認した．

# Study of Automatic GPU Offloading Technology for Open IoT

Yoji YAMATO†, Tatsuya DEMIZU†, Hirofumi NOGUCHI†, and Misao KATAOKA†

† Network Service Systems Laboratories, NTT Corporation, 3-9-11, Midori-cho, Musashino-shi, Tokyo
E-mail: †yamato.yoji@lab.ntt.co.jp

**Abstract** IoT technologies have been progressed. Now Open IoT concept has attracted attentions which achieve various IoT services by integrating horizontal separated devices and services. For Open IoT era, we have proposed the Tacit Computing technology to discover the devices with necessary data for users on demand and use them dynamically. However, existing Tacit Computing does not care about performance and operation cost. Therefore in this paper, we propose an automatic GPU offloading technology as an elementary technology of Tacit Computing which uses Genetic Algorithm to extract appropriate offload loop statements to improve performances. We evaluate a C/C++ matrix manipulation to verify effectiveness of GPU offloading and confirm more than 35 times performances within 1 hour tuning time.

## 1. はじめに

近年，IoT（Internet of Things）技術が進展しており（例えば, [1]- [18]，デバイス側で収集したデータをネットワークを介してクラウド技術（例えば, [19]- [39]）を用いて分析し可視化するといったアプリケーションが続々と出ている．従来 IoT のサービスは，デバイスからネットワーク，アプリケーションまで一体構築されたサイロ型が多かった．しかし，よりコストを下げ多様なサービスを提供するため，デバイスを複数アプリケーションで共有し，クラウド，ネットワーク，デバイスのリソースをダイナミックに連携してサービス化する Open IoT の概念が注目されている．

Open IoT では，街中の複数団体が持つ監視カメラを共有し，迷子の探索やテロリストの発見等，複数の用途に使うことが期待される．しかし，左記の例で，カメラ映像の画像処理を複数の用途で用いることは，デバイス側，クラウド側のどこで分析するとしても，CPU 計算リソースが膨大になる課題がある．

一方，近年，IoT 等多彩な分野に対応するため，CPU 以外のヘテロな計算リソースを用いることが増えている．例えば，GPU（Graphics Processing Unit）を強化したサーバで画像処理を行ったり，FPGA（Field Programmable Gate Array）で信号処理をアクセラレートすることが始まっている．Amazon Web Services (AWS) では，GPU インスタンス，FPGA インスタンスが提供されており，オンデマンドにそれらリソースを使うこともできる．Microsoft は，FPGA を用いて検索を効率化している．

Open IoT 環境では，サービス連携技術等を用いて，多彩なアプリケーションの創出が期待されるが，更に進歩したハードウェアを生かすことで，動作アプリケーションの高性能化が期待できる．しかし，そのためには，動作させるハードウェアに合

わせたプログラミングや設定が必要であり，CUDA（Compute Unified Device Architecture），OpenCL（Open Computing Language）といった多くの技術知識が求められ，ハードルは高い．

GPU や FPGA をユーザの IoT アプリケーションで容易に利用できる様にするため，動作させる画像処理，分析処理等の汎用アプリケーションを Open IoT 環境にデプロイする際に，Open IoT のプラットフォームがアプリケーションロジックを分析し，GPU，FPGA に自動で処理をオフロードすることが望まれる．

私達は，以前に Open IoT 向けのプラットフォームとして，ユーザに適切なリソースをデバイス，ネットワーク，クラウドレイヤーから発見して利用することで，ユーザにパーソナライズしたサービスを提供する Tacit Computing を提案している [40]．私達は，既存の Tacit Computing を改善し，Open IoT 環境で GPU や FPGA 等の CPU 以外の計算リソースも含めて効率的に利用することを目標としている．そのために，まず，本稿では，ユーザ向けの C/C++言語アプリケーションロジックを，自動で GPU にオフロードし，高性能化することにターゲットを絞り技術提案と評価を行う．

## 2. 既存ヘテロハードウェア処理技術

GPU の計算能力を画像処理以外にも使う GPGPU（General Purpose GPU）のための開発環境 CUDA が発展している．CUDA は GPGPU 向けの開発環境だが，GPU, FPGA, メニーコア CPU 等のヘテロハードウェアを統一的に扱うための標準規格として OpenCL も登場している．

CUDA や OpenCL では，C 言語の拡張によるプログラミングを行うが，GPU 等のデバイスと CPU の間のメモリコピー，解放等を記述する必要があり，記述の難度は高い．実際，CUDA や OpenCL を使いこなせる技術者は数多くはない．

簡易に GPGPU を行うため，ディレクティブベースで，ループ文等の並列処理すべき個所を指定し，ディレクティブに従いコンパイラがデバイス向けコードに変換する技術が有る．技術仕様として，OpenACC [41] 等，コンパイラとして PGI コンパイラ [42] 等がある．例えば，OpenACC を使った例では，ユーザは C/C++/Fortran 言語で書かれたコードに，OpenACC ディレクティブで並列処理させる等を指定する．PGI コンパイラは，コードの並列可能性をチェックして，GPU 用，CPU 用実行バイナリを生成し，実行モジュール化する．これらの技術を用いることで，GPU メモリへのデータ割り当て等を，プログラマーは意識する必要がない．

このように，OpenCL, CUDA, OpenACC 等の技術により，GPU へのオフロード処理が可能になっている．しかし，オフロード処理自体は行えるようになっても，適切なオフロードには課題が多い．例えば，Intel コンパイラの様に自動並列化機能を持つコンパイラがある．自動並列化する際は，プログラム上の for 文等の並列可能部を抽出するが，GPU を用いて並列に動作させる場合は，CPU-GPU メモリ間のデータやり取りオーバヘッドのため性能が出ないことも多い．GPU を用いて高速化する際は，スキル保持者が，OpenCL や CUDA でのチューニングや，PGI コンパイラ等で適切な並列処理部を探索することが必要になっている．[43] は，for 文が少ないベンチマークアプリケーションにて，各 for 文に対して，総当たりで並列処理するか否かを試行して性能測定を行い，最適な並列処理部の探索を行っている．

このため，スキルが無いユーザが GPU を使ってアプリケーションを高性能化することは難しいし，自動並列化技術を使う場合も，for 文の並列可否の試行錯誤等，利用開始までに多くの時間がかかっている．

## 3. Tacit Computing と GPU 自動オフロード

### 3.1 Tacit Computing 概説

Tacit Computing は，Open IoT 環境で，クラウドレイヤー，ネットワークレイヤー，デバイスレイヤーの 3 層から，ユーザにその時適切なリソースを発見，連携することで，ユーザにパーソナライズしたサービスを実現する技術である (図 1) [40]．3 層の構成であるが，時々刻々変化する状況に対応するため，ユーザに近いデバイスレイヤーでできるだけ処理を行うのがコンセプトである．Tacit Computing は，その時にユーザにあったデバイスを発見し，利用するために，ライブデータ検索技術とデバイス仮想化技術の二つを備えている．

ライブデータ検索技術は，ユーザにとって必要なデータを提供するデバイスを検索するための技術である．IoT サービスとして，定点のカメラに映った人物に対して，情報案内や警告アラート等を行うような例が考えられる．この場合は，数秒しかカメラに人物は映らないし，また，そのカメラの映像しかその人物には意味がないデータと言える．そこで，ユーザにとって必要なライブデータを検索するため，クラウドレイヤーにデータが上がるのを待つのではなく，下位レイヤーに分析する機能を配信することを，Tacit Computing では行う．例えば，マラソン大会に友人が出ており，友人が映ったカメラの映像を自動で繋いで欲しいとする．この場合，友人のゼッケン番号を検索キーにリクエストをすると，Tacit Computing では，カメラを収容するゲートウェイやネットワークエッジに，OpenCV 等の画像を分析する機能を配信して，カメラに近い場所で映像を分析することで，友人のゼッケン番号が画像分析で抽出され，友人が映っているカメラを特定する．

次に，利用したいデバイスが特定された場合に，そのデバイスを利用する必要がある．IoT デバイスは，多数のメーカーが開発しており，利用時のプロトコルやインタフェース，アドレス等が，デバイス毎に異なる．そこで，デバイス仮想化技術によって，個々のデバイスのインタフェース等の違いを吸収している．例えば，上記の例であれば，アクセス方法はカメラ毎に異なるが，Tacit Computing では，カメラ映像の取得のような共通的なリクエストを元に，カメラを収容するゲートウェイ等で，デバイス毎のアダプタにて具体的アクセス手段に変換を行い，個々のカメラに応じたリクエストを行う．デバイス仮想化技術には，Semantic Web の RDF/OWL を用いて抽象化する，Semantic Web Services 技術も利用できる (例えば，[44]- [87])．

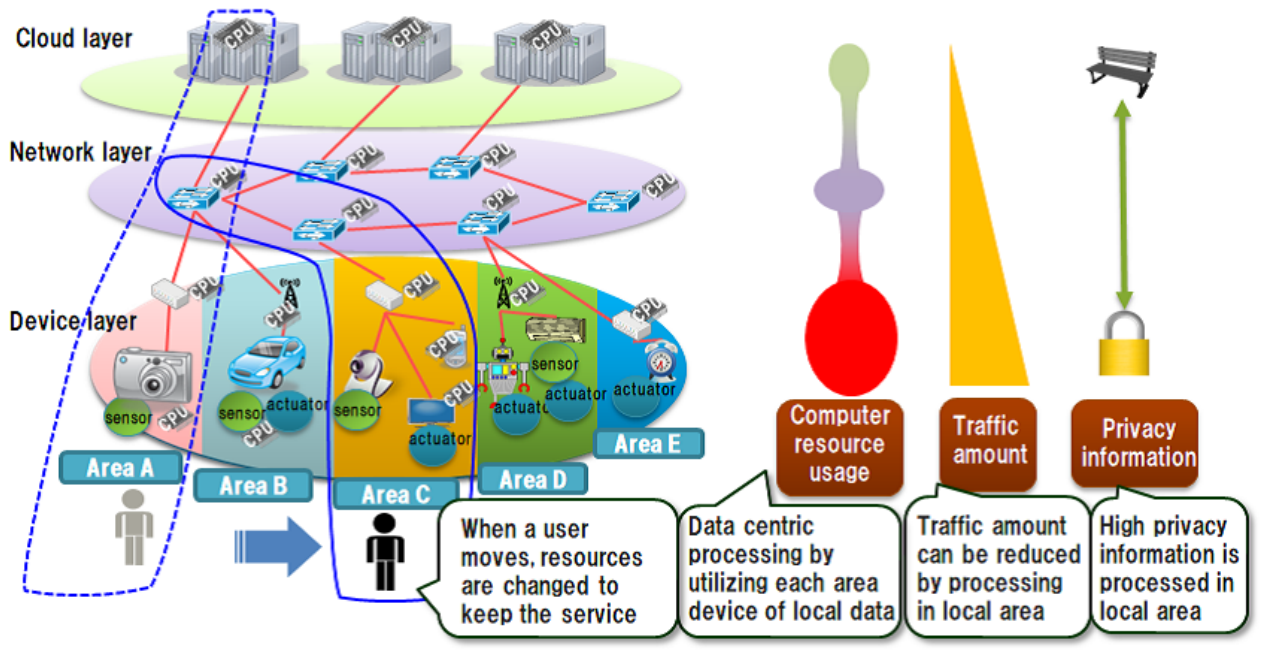

図 1　Tacit Computing 概要

### 3.2　Tacit Computing での処理オフロードによる効率化

このように，Tacit Computing により，ユーザに適切なデバイスを発見し，利用する，Open IoT のコンセプトを一部実現している．しかし，Tacit Computing で即興的にデバイスを利用，連携する場合は，コスト等は度外視されている．例えば，上記の例が，マラソン大会ランナーのモニタでなく，街中カメラを使ったテロリストの監視や高齢者の見守りだった場合は，カメラ映像を画像分析するサービスを，継続的にリーズナブルに提供する事が求められる．

そこで，私達は Tacit Computing の次のステップとして，デバイス，ネットワーク，クラウドのレイヤーで，処理オフロードを適切に行うことによる，効率化を検討している．

機能処理のオフロードのため，Tacit Computing は，ユーザが利用するアプリケーションのソースコードから，オフロードする領域を抽出して中間言語を出力し，中間言語から導かれる実行ファイルを，検証用マシンに配置実行し，オフロード効果を検証する．検証を繰り返し，適切なオフロード領域を定めたのち，Tacit Computing は，実際にユーザに提供する本番環境に，実行ファイルをデプロイし，サービスとして提供する．

図 2 を用いて，そのステップを説明する．1-1 で，Tacit Computing は，ユーザに提供しているサービスの処理機能（画像分析等）を特定する．1-2 で，Tacit Computing は，処理機能のソースコードを分析し，ループ文や FFT (Fast Fourier Transformation) ライブラリ呼び出し等の構造を把握する．1-3 で，Tacit Computing は，ループ文，FFT 等，GPU，FPGA にオフロード可能な処理を特定し，オフロード処理に応じた中間言語を抽出する．1-4 で，中間言語ファイルを出力する．なお，中間言語抽出は一度で終わりでなく，適切なオフロード領域探索のため，実行を試行して最適化するため反復される．

次に，Tacit Computing は，検証用環境として，GPU･FPGA を備えた検証用マシンに，2-1 で，中間言語から導かれる実行ファイルをデプロイする．2-2 で，配置したファイルを実行し，オフロードした際の性能を測定する．詳細は，3.3 節で後述するが，この性能測定結果を用いて，オフロードする領域をより適切にするため，1-3 の中間言語抽出のステップに戻り，別パターンの抽出を行い，性能測定を試行する．2-3 で，最終的なオフロード領域を指定したパターンを決定し，ユーザ向けの本番環境にデプロイされる．2-4 で，実行ファイル配置後，ユーザに性能を示すため，性能試験項目をテストケース DB から抽出し，抽出した性能試験を自動実施する．2-5 で，その性能試験結果を踏まえた，価格，性能等の情報をユーザに提示し，ユーザは IoT サービスの課金利用開始を判断する．

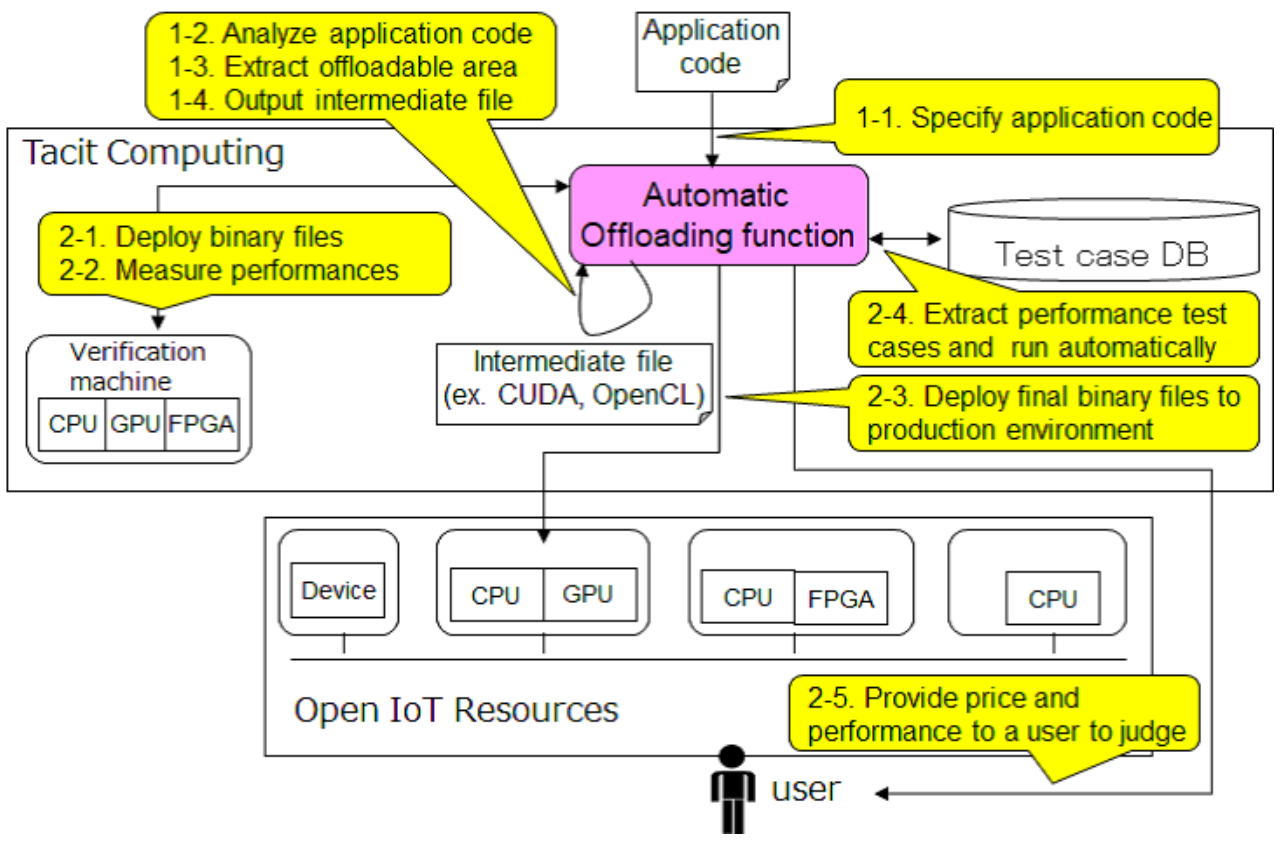

図 2　オフロード処理ステップ

### 3.3　GA を用いた GPU 自動オフロード技術の提案

本サブ節では，Tacit Computing の要素技術として，ユーザアプリケーションロジックの，GPU 自動オフロード技術を提案する．GPU 自動オフロードは，GPU に対して前サブ節の 1-3〜2-2 のステップを繰り返し，最終的に 2-3 でデプロイするオフロードコードを得るための処理である．

GPU は，一般的にレイテンシーは保証しないが，並列処理によりスループットを高める事に向いたデバイスである．IoT で動作させるアプリケーションは，多種多様であるが，カメラ映像分析のための画像処理，大量センサデータ分析のための機械学習処理等が代表的であり，それらは，繰り返し処理が多い．そこで，アプリケーションの繰り返し文を GPU に自動でオフロードする事での高速化を狙う．

しかし，2 節で記載の通り，高速化には適切な並列処理が必要である．特に，GPU を使う場合は，CPU と GPU 間のメモリ転送のため，データサイズやループ回数が多くないと性能が出ないことが多い．また，メモリデータ転送のタイミング等により，並列高速化できる個々のループ文の組み合わせが，最速とならない場合等がある．適切な並列領域指定のため，PGI コンパイラを用いて，for 文の並列可否を試行錯誤して最適化する試みがある．しかし，試行錯誤には多くの稼働がかかり，IoT サービスとして提供する際に，ユーザの利用開始が遅くなり，コストも上がってしまう問題がある．

そこで，本稿では，並列化を想定していない汎用プログラムから，自動で適切なオフロード領域を抽出するため，最初に並列可能 for 文のチェックを行い，次に並列可能 for 文群に対して GA を用いて検証環境で性能検証試行を反復し適切な領域を探索すること，を提案する．並列可能 for 文に絞った上で，遺伝子の部分の形で，高速化可能な並列処理パターンを保持し組み換えていくことで，取り得る膨大な並列処理パターンから，効率的に高速化可能なパターンを探索できると考える．

GA は，生物の進化過程を模倣した組合せ最適化手法の一つ

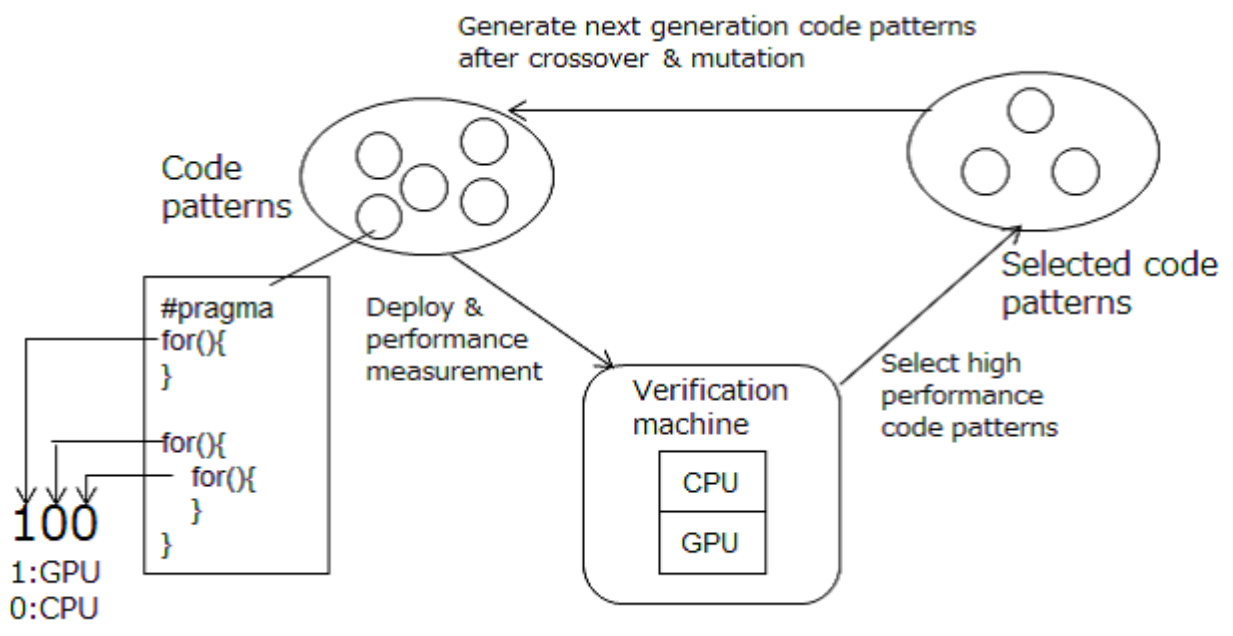

図 3　GA による GPU オフロード部の探索イメージ

である．GA のフローチャートは，初期化→評価→選択→交叉→突然変異→終了判定となっている．提案技術は，GA の中で，Simple GA を用いる．Simple GA は，遺伝子は 1, 0 のみとし，ルーレット選択，一点交叉，突然変異は 1 か所の遺伝子の値を逆にする等，単純化された GA である．

初期化では，アプリケーションコードの全 for 文の並列可否をチェック後，並列可能 for 文を遺伝子配列にマッピングする．GPU 処理する場合は 1，しない場合は 0 とする．遺伝子は指定の個体数 M が準備されるが，ランダムに 1, 0 の割り当てをする．評価では，遺伝子に該当するコードをコンパイルして検証用マシンにデプロイして実行し，ベンチマーク性能測定を行う．性能が良いパターンの遺伝子の適合度を高くする．選択では，適合度に基づいて，高適合度の遺伝子を，指定の個体数選択する．提案技術では，適合度に応じたルーレット選択及び最高適合度遺伝子のエリート選択を行う．交叉では，一定の交叉率 Pc で，選択された個体間で一部の遺伝子をある一点で交換し，子の個体を作成する．突然変異では，一定の突然変異率 Pm で，個体の遺伝子の各値を 0 から 1 または 1 から 0 に変更する．終了判定では，指定の世代数 T 回，繰り返しを行った後に処理を終了し，最高適合度の遺伝子を解とする．これら処理のイメージを図 3 に示す．

## 4. 実　　装

3 節提案技術の有効性を確認するための実装を説明する．本稿では，GA による GPU 自動オフロードの有効性確認が目的であるため，対象アプリケーションは C/C++言語のアプリケーションとし，GPU 処理自体は市中の PGI コンパイラを用いる．

C/C++言語は，OSS 及び proprietary ソフトウェアの開発で，上位の人気を誇り，数多くのアプリケーションが C/C++で開発されている．科学技術計算向けでなく，一般ユーザが用いるアプリケーションのオフロードを確認するため，分析処理や画像処理等の OSS の汎用アプリケーションを利用する．

GPU 処理は，PGI コンパイラにより行う．PGI コンパイラは OpenACC を解釈する C/C++/Fortran 向けコンパイラであり，for 文等の並列処理可能処理部を，OpenACC のディレクティブ #pragma acc kernels で指定することにより，GPU 向けバイトコードを抽出し，実行により GPU オフロードを可能としている．更に，for 文内のデータ同士に依存性があり並列処理できない処理やネストの for 文の異なる複数の階層を指定されている場合等の際に，エラーを出す．

実装の動作概要を説明する．実装は Perl 5 で行い，以下の処理を行う．処理を開始する前に，高速化する C/C++アプリケーションとそれを性能測定するベンチマークツールを準備する．

実装は，C/C++アプリケーションの利用依頼があると，まず，C/C++アプリケーションのコードを解析して，for 文を発見し，カウントする．

CPU 向け汎用アプリケーションは，並列化を想定して実装されているわけではない．そのため，まず，GPU 処理自体が不可な for 文は排除する必要がある．そこで，各 for 文一つずつに対して，並列処理の#pragma acc kernels ディレクティブ挿入を試行し，コンパイル時にエラーが出るかの判定を行う．コンパイルエラーに関しては，幾つかの種類がある．for 文の中で外部ルーチンが呼ばれている場合，ネスト for 文で異なる階層が重複指定さている場合，break 等で for 文を途中で抜ける処理がある場合，for 文のデータにデータ依存性がある場合等がある．アプリケーションによって，コンパイル時エラーの種類は多彩であり，これ以外の場合もあるが，コンパイルエラーは処理対象外とし，#pragma ディレクティブは挿入しない．

ここで，並列処理してもエラーが出ないループ文の数が a の場合，a が遺伝子長となる．遺伝子の 1 は並列処理ディレクティブ有，0 は無に対応させ，長さ a の遺伝子に，アプリケーションコードをマッピングする．

次に，初期値として，指定個体数の遺伝子配列を準備する．0 と 1 をランダムに割当てて作成する．

準備された遺伝子配列に応じて，遺伝子の値が 1 の場合は並列処理を指定するディレクティブを C/C++コードに挿入する．作成した C/C++コードを，GPU を備えたマシン上の PGI コンパイラでコンパイルを行う．コンパイルした実行ファイルをデプロイし，ベンチマークツールで性能を測定する．

全個体数に対して，ベンチマーク性能測定後，ベンチマーク処理時間に応じて，各遺伝子配列の適合度を設定する．設定された適合度に応じて，残す個体の選択を行う．選択された個体に対して，交叉処理，突然変異処理，そのままコピー処理の GA 処理を行い，次世代の個体群を作成する．

次世代の個体に対して，コンパイル，性能測定，適合度設定，選択，交叉，突然変異処理を行う．ここで，GA 処理の中で，以前と同じパターンの遺伝子が生じた場合は，その個体についてはコンパイル，性能測定をせず，以前と同じ測定値を用いる．指定世代数の GA 処理終了後，最高性能の遺伝子配列に該当する，C/C++コードを解とする．

この中で，個体数，世代数，交叉率，突然変異率，適合度設定，選択方法は，GA のパラメータであり，評価時の 5 節で指定する．

## 5. 評　　価

### 5.1 評 価 条 件

#### 5.1.1 評 価 対 象

評価対象は，IoT で多くのユーザが利用すると想定される行

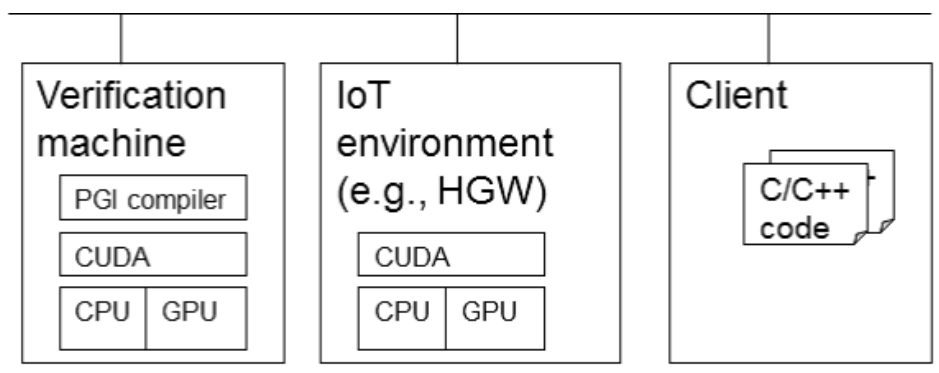

| Name | Hardware | CPU | RAM | GPU | OS | CUDA toolkit | PGI compiler |
|---|---|---|---|---|---|---|---|
| Verification Machine | Dell New Vostro 15 5000 | Intel Xeon E5-2620 v3@2.40GHz | 32GB | NVIDIA Quadro K5200 (CUDA core: 2304, Memory: GDDR5 8GB) | Ubuntu 16.04.3 | 9.1 | 17.1 |
| IoT environment (e.g., HGW) | Dell New Vostro 15 5000 | Intel Xeon E5-2620 v3@2.40GHz | 32GB | NVIDIA Quadro K5200 (CUDA core: 2304, Memory: GDDR5 8GB) | Ubuntu 16.04.3 | 9.1 | |
| Client | Lenobo ThinkPad X1 carbon | Intel Core i5 5200U@2.20GHz | 8GB | | Windows 7 Professional | | |

図 4　性能測定環境

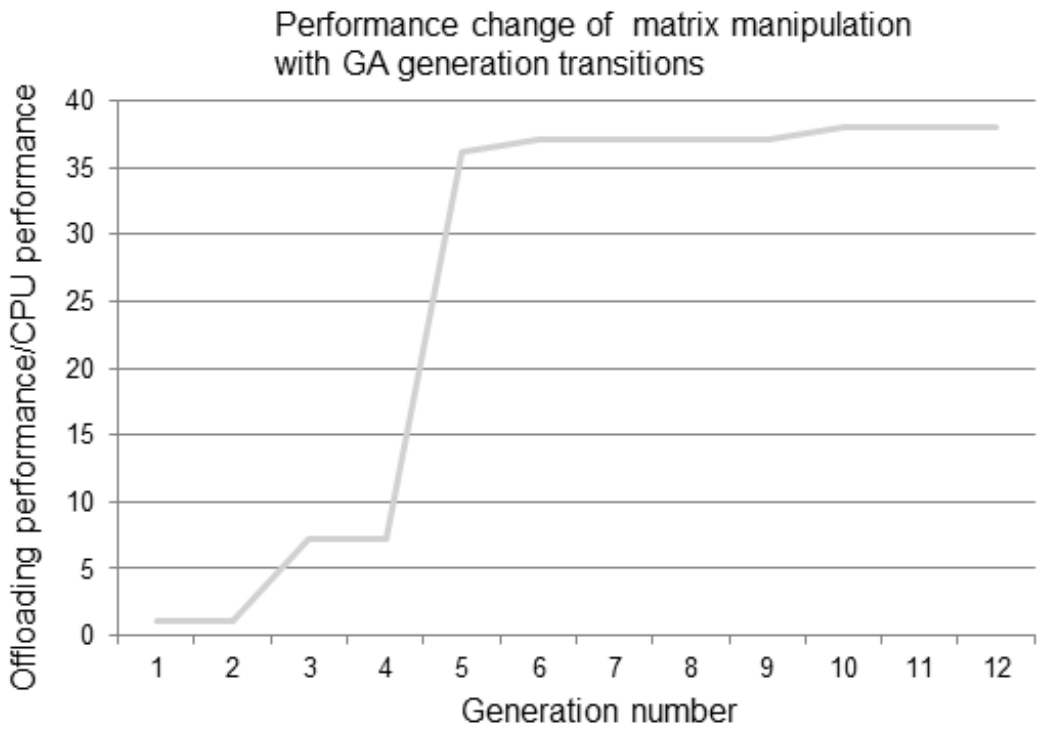

図 5　GA 世代数に伴う行列処理の性能変化

列計算とする．

行列積計算は，機械学習分析の様々な場面で利用されている．Open IoT で，デバイスからデータをネットワークで転送するアプリケーションを考えた際に，ネットワークコストを下げるため，デバイス側で行列計算等の一次分析を送ることは考えられる．チューニングするためのベンチマークとしては，2048 × 2048 の行列積を計算するテストを行うツールを利用する．

**5.1.2 評価手法**

GA の世代変化を通じて，ベンチマーク性能が変化することを確認する．

実行する Simple GA の，パラメータ，条件は以下で行う．

遺伝子長：並列可能ループ文数（行列計算は 12）

個体数 M：遺伝子長以下とする（行列計算は 12）

世代数 T：遺伝子長以下とする（行列計算は 12）

適合度：(ベンチマーク処理時間)$^{-1/2}$

選択：ルーレット選択．ただし，世代での最高適合度遺伝子は交叉も突然変異もせず次世代に保存するエリート保存も合わせて行う．

交叉率 Pc：0.9

突然変異率 Pm：0.05

**5.1.3 評価環境**

利用する GPU として NVIDIA Quadro K5200 を備えた物理マシンを検証に用いる．NVIDIA Quadro K5200 の CUDA コア数は 2304 である．PGI コンパイラはコミュニティ版の 17.10，CUDA Toolkit は 9.1 を用いる．評価環境とスペックを図 4 に示す．

### 5.2 性能結果

図 5 に，行列処理の，各世代個体の最高性能と GA の世代数をグラフにとる．性能は CPU のみで処理の場合との比で示している．図 5 より，12 世代の GA の中で，性能が向上しているのが分かり，全て 0（全 CPU 処理）の遺伝子では 92.27 秒だったのが，12 世代目で 2.43 秒で処理し，37 倍の性能が実現出来ていることが分かる．また，GA の各遺伝子評価の一回の試行はタイムアウト含めて平均 2 分弱であり，12 世代トータルで最大数時間程度かかるはずだが，同パターンの処理スキップのおかげで，1 時間以内でオフロード抽出処理は終わった．

私達は，早期に本番サービスが開始できるようにするため，GA の個体数，世代数とも絞っている．そのため，性能向上が収束してはいないことは明確であるが，20 世代でも CPU に比べて十分な性能が得られていることが分かる．

## 6. ま と め

本稿では，Open IoT 環境で，GPU をユーザアプリケーションで有効活用するための，Open IoT プラットフォームの要素技術として，GPU 自動オフロード技術を提案した．

提案技術は，デプロイするアプリケーションのコードから，上位のループ文を検出し，それらの並列化可否を，実際のベンチマーク性能結果に基づいた GA で試行探索し，適切なオフロード部を抽出する．C/C++の OSS アプリケーションを対象に，PGI コンパイラを用いて提案技術を実装した．評価では，行列計算に対して，12 世代，1 時間以内の短時間の試行で，CPU 処理に比べて 35 倍以上の高性能化ができることを確認した．

今後は，GA のパラメータ検討，適用範囲拡大により，より短時間，高性能でのオフロードの探索を検討する．

## 文　献